\documentclass[a4paper]{article}

\usepackage{INTERSPEECH_v2}

\title{Combining Spatial Clustering with LSTM Speech Models for Multichannel Speech Enhancement}
\name{Felix Grezes$^1$, Zhaoheng Ni$^1$, Viet Anh Trinh$^1$, Michael Mandel$^{12}$}
\address{
  $^1$The Graduate Center, City University of New York, USA\\
  $^2$Brooklyn College, City University of New York, USA}
\email{fgrezes@gradcenter.cuny.edu, zni@gradcenter.cuny.edu,  vtrinh@gradcenter.cuny.edu, mim@mr-pc.org}

\begin{document}

\maketitle

\begin{abstract}
%
%
Recurrent neural networks using the LSTM architecture can achieve significant single-channel noise reduction. It is not obvious, however, how to apply them to multi-channel inputs in a way that can generalize to new microphone configurations. In contrast, spatial clustering techniques can achieve such generalization, but lack a strong signal model.

This paper combines the two approaches to attain both the spatial separation performance and generality of multichannel spatial clustering and the signal modeling performance of multiple parallel single-channel LSTM speech enhancers. 

The system is compared to several baselines on the CHiME3 dataset in terms of speech quality predicted by the PESQ algorithm and word error rate of a recognizer trained on mis-matched conditions, in order to focus on generalization. 

Our experiments show that by combining the LSTM models with the spatial clustering, we reduce word error rate by 4.6\% absolute (17.2\% relative) on the development set and 11.2\% absolute (25.5\% relative) on test set compared with spatial clustering system, and reduce by 10.75\% (32.72\% relative) on development set and 6.12\% absolute (15.76\% relative) on test data compared with LSTM model.

\end{abstract}


\section{Introduction}
With speech recognition techniques approaching human performance on noise-free audio with a close-talking microphone \cite{achieving-human-parity-conversational-speech-recognition-2}, recent research has focused on the more difficult task of speech recognition in far-field, noisy environments. 

One solution is beamforming, which combines multiple microphone channels in order to optimize some criterion \cite{brandstein2013microphone}, which is typically based on properties of the signals or the spatial configuration of the recordings  at test time, with no training ahead of time.

Recurrent neural networks using the LSTM architecture can achieve significant single-channel noise reduction \cite{6638947, weninger2015speech}, and so there is interest in using trainable deep-learning models to perform beamforming. This is especially useful for optimizing beamformers directly for performing automatic speech recognition \cite{xiao16, SainathEtAl2015}, although such optimization must happen at training time on a corpus of training data.  Such models have difficulty generalizing across microphone arrays, including differences in number of microphones and array geometries, such as the AMI corpus \cite{carletta2007unleashing,4430116} and the CHIME challenge \cite{barker2015third}.  This is similarly reflected in difficulties generalization of certain types of beamforming across conditions \cite{vincent2016analysis}.

In contrast to deep learning-based beamforming, spatial clustering is an unsupervised method for performing source separation, so it easily adapts across microphone arrays \cite{5200357, sawada2011underdetermined, bagchi15}.  Such methods group spectrogram points based on similarities in spatial properties, but are typically not able to take advantage of signal models, such as models of speech or noise. 
Developed by Mandel et al \cite{5200357}, Model-based EM Source Separation and Localization (MESSL) is a system that computes spectrogram masks for source separation as a byproduct of estimating the spatial location of the sources. It does so using the EM (expectation maximization) algorithm, iteratively refining the estimates of the spatial parameters of the audio sources and the spectrogram regions dominated by each source.

The goal of this paper is to augment the capabilities of MESSL using neural network trained speech signal models.

Other recent approaches to combine beamforming with ASR to optimize the beamformer for metrics on the ASR output \cite{heymann2016neural,heymann2017}. 


%

%


In this paper we describe several methods for incorporating single-channel LSTM-based speech enhancement into MESSL to improve its performance.

The first approach is a naive combination of the parallel single-channel LSTM models \cite{erdogan2016improved} with the MESSL mask.  It is able to increase PESQ score from 1.57 to 2.49 (see Table \ref{tab:pesq_real}), and reduce WER from 43.94\% to 36.05\% (See Table \ref{tab:wer_real}) on the CHIME3 evaluation dataset of real noisy environmental audio.
The second is to use the LSTM-based mask to initialize MESSL.  This provides a comparable PESQ score of 2.46, while further improving the WER to 32.72\% . (See Tables \ref{tab:pesq_real} and \ref{tab:wer_real}).

\section{Related Work}

Recently, Nugraha et al. \cite{nugraha2016multichannel} also studied multi-channel source separation using deep feedforward neural networks, using a multi-channel Gaussian model to combine the source spectrograms, to take advantage of the spatial information present in the microphone array. They explore the efficacy of different loss functions and other model hyper-parameters. One of their findings is that the standard mean-square error loss function performed close to the best.

Pfeifenberger et al. \cite{pfeifenberger2017} proposed an optimal multi-channel filter which relies solely on speech presence probability. This speech-noise mask is learned using a 2-layer feedforward neural network, trained on the simulated noisy data portion of CHIME-3. They show that this filter improves the PESQ score of the audio.

Heymann et al \cite{heymann2016neural,heymann2017} also study the combination of multi-channel beamforming with single-channel neural network model. Similar to ours, the proposed model consists of a bidirectional LSTM layer, followed by feedforward layers, in their case three. Of particular note is the companion paper by Boeddeker et al. \cite{boeddeker2017}, which provides a major theoretical breakthrough by generalizing the backpropagation algorithm to complex values. This allows their system to run on complex spectrograms, and not real valued approximations.

\section{Methods}

\subsection{The CHIME-3 Corpus}
The CHIME-3 corpus features both live and simulated, 6-channel single speaker recordings from 12 different speakers (6 male, 6 female), in 4 different noisy environments: caf\'{e}, street junction, public transport and pedestrian area. 
In our work, we used the proposed data split, with 1600 real noisy utterances in the training set for training, 1640 real noisy utterances in the development set for validation. We did not use the simulated data to train our models. We tested our models on the proposed 2640 utterances in the test set, which contains audio both from real noisy recordings and simulated noisy recordings.
In order to perform speech recognition, we used the Kaldi \cite{Povey_ASRU2011} toolkit trained on the AMI corpus, which features 8 microphones, recording overlapping speech in meeting rooms. 
These differences provide an additional challenge, but are essential to training a robust model and testing it in real-world conditions.




\subsection{Baseline: Online Multichannel Noise Tracking and Reduction}
Minima tracking and minima controlled recursive averaging (MCRA) are two dominant approaches to estimate single channel noise power.
The minima tracking method \cite{martin2001noise} is used to estimate the power of the noise signal. This technique is based on the observation that the power of a noisy speech signal is the same as the power of only the noise signal when speech is absent, even during an utterance. Therefore, the noise power can be estimated by tracking the minimum of the noisy signal in a proper number of time frames of a periodogram.

In minima controlled recursive averaging (MCRA) \cite{cohen2003noise} and improved minima controlled recursive averaging (IMCRA) \cite{cohen2002noise}, the noise measurement is recursively estimated based on past and current frame information when speech is absent, but the measurement is held when speech is present. An estimated speech presence probability is required to identify the absence and presence of speech.

Souden et al. \cite{souden2011integrated} applied these single-channel noise power estimation methods to computing multi-channel noise covariance matrix using a multi-channel speech presence probability (MC-SPP) estimate.  We implemented this approach ourselves and use it as a baseline in our experiments.
These methods rely on purely statistical information in the data to estimate the noise and speech probability distributions, and do not use spatial clustering or machine learning.

\subsection{Supervised MVDR Speech Reference}

Because the real subset of the CHiME-3 recordings were spoken in a noisy environment, it is not possible to provide a true clean reference signal for them.  Instead, an additional microphone was placed close to the talker's mouth to serve as a reference.  While this reference has a higher signal-to-noise ratio than the main microphones, it is not noise free.  In addition, because it is mounted close to the mouth, it contains sounds that are not desired in a clean output and actually could hurt ASR performance, namely pops, lip smacks, and other mouth noises.

In order to obtain a cleaner reference signal, we use the close microphone as a frequency-dependent voice activity detector to control mask-based MVDR beamforming \cite{souden2010} of the main six microphones.  The beamformer is adapted in a given frequency when the reference microphone energy is greater than the 85th percentile in that frequency band over an entire utterance.  A final post-filter mask is derived in a similar way, but using the 75th percentile, and then applied with a maximum suppression of 15\,dB.  By not including the close microphone signal directly in the estimate, we eliminate the recording artifacts that it contains, while still taking advantage of the information on voice activity that it contains.  This is in contrast to other approaches to estimate a reference signal \cite{vincent2016analysis}, which learn a linear transformation to project the close microphone signal onto the microphone array. Such an approach can compensate for different filtering of the speech signal at the close mic and the main mic, but cannot eliminate recording artifacts.

\subsection{Training the LSTM Speech Signal Model}
Using the above reference signal for all microphone channels, we compute ideal amplitude masks \cite{erdogan2015phase}:
\[ m_{ia}(\omega,t) =  |s(\omega,t)|/|y(\omega,t)|\]
with $s(\omega,t),y(\omega,t)$ being the complex spectrograms of the reference and noisy microphone channel respectively, obtained by short-time  Fourier transform. We then trained an LSTM neural network which takes as input a time series of noisy spectrogram frames, and outputs for each frame a $[0,1]$ valued mask $\hat{m}(\omega,t)$, trained to minimize the mean-squared error
\begin{align}
\mathcal{L}(\Theta) = \sum_{\omega,t} \left( \hat{m}(\omega,t) |y(\omega,t)| - |s(\omega,t)| \right)^2,
\end{align}
where $\Theta$ represents the model parameters of the LSTM. Ideally $\hat{m}(\omega,t)$ should closely match $m_{ia}(\omega,t)$, especially in points of high-energy in the spectrogram.

We tried various sizes and numbers of LSTM layers. The spectrogram input was converted from a linear to decibel scale, and normalized to mean 0 and variance 1 at each frequency. To perform the computation and training of our LSTM neural network, we used the KERAS python library \cite{chollet2015keras}, built upon the Tensorflow library \cite{tensorflow2015-whitepaper}.

After extensive hyper-parameter exploration, we found the following network parameters to produces the best results:
\begin{itemize}
\item Single bidirectional LSTM layer of size 1024, with the forward and backward activations being concatenated.
\item RMSprop for the batch gradient back-propagation optimizer, with default parameters.
\item Batch size for learning of 128 sequences of length 1600ms (50 spectrogram frames at 1024 overlapping window size).
\end{itemize}
The model was trained until the loss on the validation set no longer improved. 

\begin{figure}[h!]
    \centering
    \includegraphics[width=0.5\textwidth]{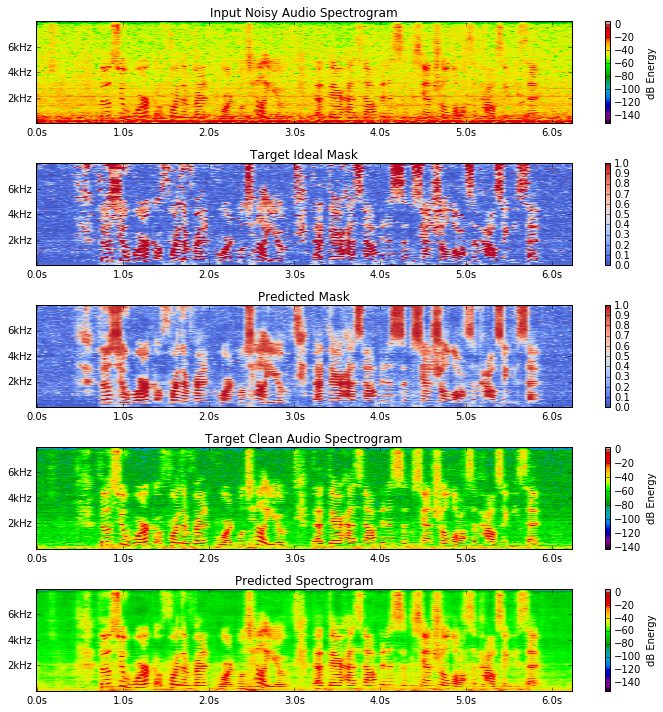}
    \caption{An example of the input, target, and output of the LSTM model along with the predicted spectrogram which is derived by applying the predicted mask to the input.}
    \label{fig:ex1}
\end{figure}

Figure \ref{fig:ex1} gives a visual example of how our model enhances specific areas related to speech of a spectrogram. Results for this model enhancing speech over the development set and  test set are shown in Tables \ref{tab:pesq_simu},\ref{tab:pesq_real}, and \ref{tab:wer_real}.
 
\begin{figure*}[h!]
    \centering
    \includegraphics[width=0.95\textwidth]{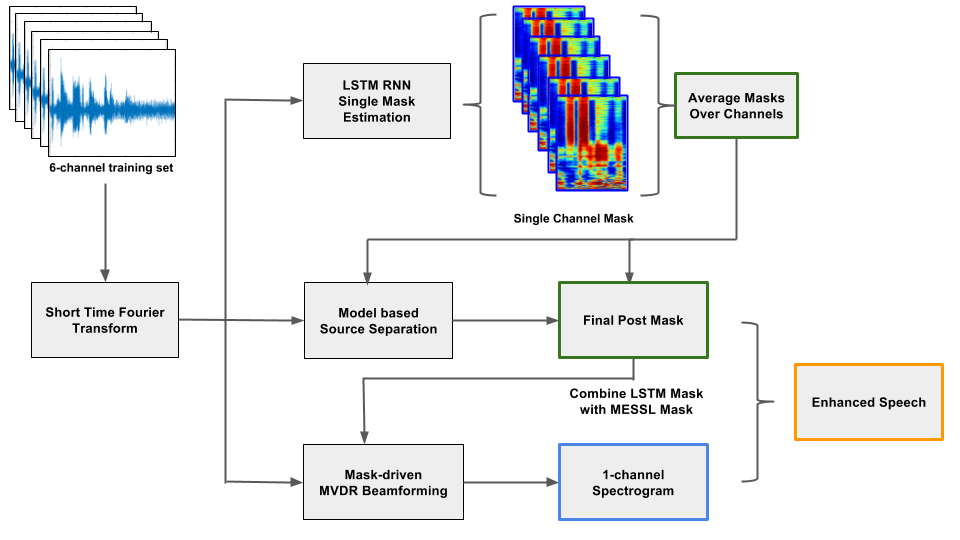}
    \caption{Multi-channel Speech Enhancement System}
    \label{fig:framework}
\end{figure*}

\newpage

\section{Experiments}
%
%

We report the performance of several experiments which we compare to previous works, shown in the Tables \ref{tab:pesq_simu},\ref{tab:pesq_real} and \ref{tab:wer_real}.
\begin{itemize}
\item LSTM: Applies the mask generated by the LSTM model described in section 3.4.
\item MESSL: Applies the mask generated by MESSL.
\item MESSL+LSTM: Combines the masks generated by MESSL and the LSTM model using different combination types (average, max, min).
\item LSTM-initial MESSL: Uses LSTM mask as the initial mask for MESSL's EM optimization.
\item Channel-5, Souden et al., Erdogan et al. are described in \cite{barker2015third, souden2011integrated,erdogan2016improved} respectively.
\end{itemize}

A flowchart illustrating the processing in our basic system is shown in Figure~\ref{fig:framework}.
We extract six spectrograms from six-channel audio files using a window size is 1024 (64ms at 16kHz). We use the LSTM model to predict the mask for each spectrogram, and take the average of the six predicted masks to obtain a single mask. MESSL also estimates a single mask from multi-channel inputs. We first try to combine the two masks directly by taking the maximum, average, and minimum of the masks at each time-frequency point. Then we perform mask-driven MVDR beamforming using the combined mask. We apply different combinations of masks onto the corresponding enhanced spectrogram and get the enhance audio using the Inversed Short-time Fourier Transform. 

Since MESSL is based on the EM algorithm, it is important to initialize it well. We use LSTM averaged mask as the initial mask for MESSL approach, after one EM iteration, average the output with LSTM mask and use it for next iteration, we do the same process for the first few iterations and let MESSL finish. Then we also average the LSTM mask with MESSL mask as the driven mask for both MVDR beamforming and the post mask for speech enhancement. We find holding LSTM mask for 11 iterations gets the best result. The results achieve the state-of-art on simulate set and real set. In this system We can combine the spatial information obtained by MESSL with single channel separation result from LSTM model and get a cleaner mask. 

We use two different metrics to compare the performance of these systems. The first metric is the Perceptual Evaluation of Speech Quality (PESQ) \cite{rix2001perceptual}, which evaluates the speech quality relative to a reference.  We evaluate these algorithms on both the simulated and real portions of the CHiME3 test set using the PESQ measure.  The second metric is word error rate (WER) of a mismatched recognizer, intended to measure the ability of the model to permit generalization to new datasets. The ASR is trained from the AMI corpus training set \cite{carletta2007unleashing,4430116} processed using the BeamformIt algorithm \cite{Anguera-IEEE-07}. The purpose of using the AMI-trained recognizer is to measure how much we can reduce the mismatch of training data and test data since the AMI corpus has little noise inside while the original CHiME3 test data is recorded in noisy environments.

\subsection{Results}

The results in Table \ref{tab:pesq_simu} show PESQ results on the simulated data.  The baseline systems, Souden, et al.\cite{souden2011integrated}, Erdogan, et al.\cite{erdogan2016improved}, MESSL, the parallel single-channel LSTM models all perform similarly. In Erdogan, et al.'s paper\cite{erdogan2016improved}, they take the maximum of six masks predicted by LSTM model. And they apply the mask with minimum floor as the post mask after MVDR beamforming, while we take the average of six masks and apply it directly as the post mask. Among the proposed combinations of MESSL and the LSTM mask, the average of the LSTM mask and the MESSL mask achieves the highest PESQ score, but the LSTM-initial MESSL model outperforms all the other results on simulated set.
%
%
PESQ results on the real set are shown in Table~\ref{tab:pesq_real}, showing that real set is more difficult to enhance.
The average of the LSTM mask and the MESSL mask outperform the baseline results, however. The LSTM-initial MESSL model achieves better PESQ score on development set but fails to beat MESSL+LSTM average model on  evaluation set. 
%
%

%
%
\begin{table}[h!]
\caption{\label{tab:pesq_simu} {\it PESQ scores for CHiME-3 simulated set}}
\vspace{2mm}
\centerline{
\begin{tabular}{| l | c | c | c |}
\hline
Models & Combination & dev & eval\\
\hline \hline
Channel-5 & None & $2.32$ & $2.35$ \\
MESSL & MESSL&  $2.47$ &       $2.44$ \\
LSTM & LSTM& $2.51$ &         $2.34$ \\
Souden, et al. &None &  $2.31$ &        $2.44$ \\
Erdogan, et al. & Minfloor & $2.19$ &  $2.29$\\
MESSL+LSTM &Average & $3.15$ & $3.13$ \\
MESSL+LSTM  & Max&  $3.13$ &        $3.10$ \\
MESSL+LSTM  &Min &  $2.82$ &        $2.87$ \\
LSTM-initial MESSL &Average &  $\underline{3.26}$ &        $\underline{3.24}$ \\
\hline
\end{tabular}
}
\end{table}

\begin{table}[h!]
\caption{\label{tab:pesq_real} {\it PESQ scores for CHiME-3 real set}}
\vspace{2mm}
\centerline{
\begin{tabular}{| l | c | c | c |}
\hline
Models & Combination & dev & eval \\
\hline \hline
Channel-5 & None & $2.02$ & $1.92$\\
MESSL & MESSL& $1.92$ &       $1.57$ \\
LSTM & LSTM& $2.51$ &         $2.42$ \\
Souden,et al. &None &  $2.14$ &        $2.05$ \\
Erdogan, et al. & Minfloor & $1.68$ & $1.79$\\
MESSL+LSTM &Average & $2.73$ & $\underline{2.49}$ \\
MESSL+LSTM  & Max& $2.67$ &        $2.43$ \\
MESSL+LSTM  &Min &  $2.38$ &    $2.10$ \\
LSTM-initial MESSL &Average & \underline{$2.76$} &        $2.46$ \\
\hline
\end{tabular}
}
\end{table}

Table \ref{tab:wer_real} shows the WER results. Since the purpose of the model is to reduce the mismatch between training and test data, we are more interested in the results on real data. The WER of the LSTM-enhanced speech is higher than that of the MESSL model. The result shows spatial clustering can get a better performance compared with using single channel mask estimation. We find using the LSTM mask as the initial mask for MESSL approach outperforms the baseline results and three kinds of mask combination results. Compared with Erdogan, et al's model \cite{erdogan2016improved}, 
we achieve 2.1\% WER reduction on the evaluation set.

Taking the minimal of LSTM mask and MESSL mask did the worst on WER, which means much information in speech frequency region has been suppressed. Taking the maximum of two masks outperforms the average mask's model on evaluation data. It means the noise in two models has been suppressed sort of equally, but how to enhance the speech regions on the spectrogram is still an interesting and challenging task.

\begin{table}[h!]
\caption{\label{tab:wer_real} {\it WER (\%) results on CHiME-3 real set using Kaldi ASR trained on AMI corpus}}
\vspace{2mm}
\centerline{
\begin{tabular}{| l | c | c | c |}
\hline
Models & Combination & dev & eval\\
\hline \hline
Channel-5 & None & $34.8$ & $55.5$\\
MESSL & MESSL& $26.6$ &       $43.9$ \\
LSTM & LSTM& $32.9$ &         $38.9$ \\
Souden, et al &None & $37.4$ &        $52.3$ \\
Erdogan, et al & Minfloor & $21.2$ & $34.8$ \\
MESSL+LSTM &Average & $22.6$ & $36.1$ \\
MESSL+LSTM  & Max& $22.8$ &        $33.8$ \\
MESSL+LSTM  &Min & $54.2$ &        $66.3$ \\
LSTM-initial MESSL &Average & $\underline{22.1}$ &        $\underline{32.7}$\\
\hline
\end{tabular}
}
\end{table}

%
%
\section{Conclusion and Future Work}
In this paper we propose a system to adapt parallel single-channel LSTM-based enhancement to multi-channel audio, combining the speech-signal modeling power of the LSTM neural network with spatial clustering power of MESSL, further enhancing the audio.

Our future work will continue to explore different ways of integrating the LSTM speech-signal model with MESSL. Preliminary results show that the spatial information is more valuable than the single-channel information with respect to WER, the next logical step is to integrate a mask cleaning LSTM model in each loop of MESSL's EM algorithm, i.e train a mask-cleaner LSTM model which takes for input both the noisy spectrogram and a mask produced by MESSL, produces a new mask, with the cleaned speech as target. 

We also would like to try phase-sensitive mask for the spectrogram approximation, as \cite{erdogan2015phase} report better results using those. Phase-sensitive masks are computed by $m_{ps}(w,t) = cos(\theta^s-\theta^y)*|s(\omega,t)|/|y(\omega,t)|$, with $\theta^s,\theta^y$ being the angles of the complex valued clean and noisy spectrograms respectively. Another idea is to work directly on complexed-valued spectrograms, neural networks and masks, as in \cite{boeddeker2017}.

\newpage

\bibliographystyle{IEEEtran}

\bibliography{mybib}


\end{document}